# Generative AI in Higher Education Laboratory Learning: A Qualitative Case Study of Epistemic Scaffolding and Assessment Boundaries

Matteo Tuveri[1,2], Alessandro Riggio[1,2,3]

[1] Physics Department, University of Cagliari, Cittadella Universitaria di Monserrato, 09042, Monserrato (CA), Italy

[2] INFN sezione di Cagliari, Cittadella Universitaria di Monserrato, 09042, Monserrato (CA), Italy

[3] INAF – IASF Palermo, Via U. La Malfa 153, 90146 Palermo, Italy

***Corresponding author**: matteo.tuveri@ca.infn.it

## Abstract
Advanced physics laboratories require students to integrate disciplinary knowledge, experimental practice and scientific argumentation across complex observational and analytical tasks. The increasing availability of generative artificial intelligence (GenAI) adds complexity to this coordination, since AI systems may function as conceptual explainers, operational assistants, artefact reviewers or apparently authoritative evaluators. This exploratory qualitative case study examines AstroTutor, a constrained GenAI tutor introduced as an optional support resource in a Master's-level advanced astrophysics laboratory. The study investigates how students framed the tutor within a broader GenAI-mediated learning ecology that included the instructor, peers, course materials, observations, measurements, data analysis and final assessed reports. Seven students attended the course, five used the tutor, and three groups produced a final report. The analysis combined content analysis, thematic analysis and frame analysis. Drawing on chat logs, final reports and limited post-use reflective responses, the results identify five principal GenAI functions: interface interpreter, warrant organiser, report scaffold, unstable authority and resource whose traces may appear in downstream reports. These findings extend previous research on GenAI in education to the context of advanced physics laboratories, showing that its use requires explicit design boundaries, guidance on legitimate and prohibited practices, verification routines, and assessment requirements that preserve students' epistemic responsibility. The educational implications of a GenAI-mediated learning ecology in advanced physics laboratories are also discussed.

Keywords: generative artificial intelligence; physics education research; advanced astrophysics laboratory; learning ecology; epistemic framing

## 1. Introduction
Generative artificial intelligence (GenAI) has rapidly entered physics education and, more broadly, higher education. Recent research shows that large language models can support explanation, feedback, tutoring, coding assistance and flexible access to learning resources [1-8]. The same literature, however, also highlights persistent concerns about inaccurate content, over-reliance, opaque reasoning, accountability, assessment practices and the need for human oversight. These issues are particularly relevant when students work on complex disciplinary artefacts in which reasoning, technical decisions, data analysis and scientific communication are closely intertwined. In such contexts, the educational value of GenAI cannot be inferred from the fluency of its responses. It depends on how AI-generated support is situated within task structure, epistemic authority, validation practices and assessment expectations.

Research on intelligent tutoring systems provides an important reference point for this problem. Educationally meaningful tutoring requires more than information delivery: classical systems model domain knowledge, learner activity, feedback and task structure, while dialogue-based tutors organise natural-language interaction around pedagogical goals [9-11]. LLM-based tutors differ from these systems because they can produce fluent and context-sensitive dialogue without stable guarantees of correctness, learner modelling or instructional alignment [12-16]. This makes constraint and role definition central design problems. A GenAI tutor may support reasoning when it prompts clarification, elicits assumptions, connects representations to decisions or provides diagnostic feedback on student-produced artefacts. It may also distort learning activity when it produces final answers, simulates evaluative authority or encourages students to treat generated prose as a substitute for disciplinary judgement.

Existing physics education research has begun to examine how students use and evaluate GenAI in relation to physics learning. Students' ability to judge AI-generated physics explanations depends on their conceptual knowledge; laboratory uses of ChatGPT require teacher guidance and critical verification; AI-based tutoring can support explanation but also create risks of over-reliance; and AI-generated language in laboratory reports raises specific concerns about authorship and assessment [5,12-14]. These studies converge on a central point: GenAI should not be analysed only as a source of answers, but as a mediating resource whose educational meaning depends on disciplinary knowledge, task structure, epistemic authority and validation practices.

Advanced university courses amplify these issues. Students often work across specialised disciplinary knowledge, digital tools, data products, collaborative activity and assessed written artefacts. GenAI can therefore enter the same space in which they plan work, interpret representations, debug procedures, formulate explanations and prepare reports. The central problem is not whether GenAI can provide useful answers, but how its role can be designed and interpreted without blurring the boundaries between scaffolding and task completion, feedback and authorship, verification and acceptance, formative support and assessment authority, with careful attention to reliability, agency, assessment and human oversight [17,18].

Advanced laboratory learning provides a demanding context for examining these tensions. Laboratory work in higher education is not reducible to recalling definitions or applying formulas; it requires students to coordinate conceptual, experimental, methodological and argumentative resources in order to produce defensible scientific claims [19-24]. In physics and astronomy laboratories, this coordination is especially visible because students must connect physical concepts, instrumental constraints, software-mediated representations, data-reduction procedures and reportable claims. Although disciplinary specificity remains important, a cross-cutting problem is emerging in higher education laboratory learning: how AI-mediated scaffolding is framed, how epistemic warrants become visible across artefacts, and how assessment boundaries are maintained or threatened.

The present study investigates AstroTutor, a constrained GenAI tutor implemented in a Master's-level astrophysics laboratory to provide always-available support without replacing the instructor. The tutor was configured around the optical photometry activity, the course tool Observation Planner, and an instructional script that limited its intended role to conceptual clarification, interface interpretation, diagnostic feedback, controlled operational support and report-oriented scaffolding. The analysis focuses on how students framed the tutor, which epistemic functions emerged in the interactions, whether warrants visible in AI-mediated exchanges were traceable in downstream artefacts, and what risks appeared when AI support entered an activity culminating in a final assessed report.

The study is exploratory and qualitative. It does not evaluate AstroTutor's effectiveness, measure learning gains or claim statistical generalisation. Its aim is to provide a theoretically informed account of GenAI-mediated scaffolding and assessment boundaries in an authentic higher education laboratory ecology. In doing so, it contributes to research on AI-mediated laboratory learning by showing how a constrained tutor can become an interface interpreter, warrant organiser, report scaffold and unstable authority, depending on students' framings and on the artefacts around which interaction is organised.

The study addresses the following research questions:

- RQ1. How do students frame a constrained AI tutor when working in an advanced astrophysics laboratory course?
- RQ2. Which epistemic warrants, including concept-to-decision links, are visible in AI-mediated interactions and traceable, absent, transformed or weakened across students' laboratory artefacts, reflective responses and final reports?

# 2. Theoretical framework

## 2.1 Advanced laboratory learning as an epistemic activity

Physics laboratory learning is increasingly understood as an epistemic activity in which students must use disciplinary knowledge to act within experimental situations, rather than merely recall definitions or apply formulas. In this perspective, concepts acquire educational value when they guide observation, support methodological choices, constrain measurement procedures, inform uncertainty evaluation and contribute to defensible scientific claims [19-24].

Epistemological framing research in physics education supports this view by showing that students' reasoning depends not only on the conceptual resources they possess, but also on how they interpret the activity, what they treat as relevant knowledge and what they accept as a legitimate warrant in a given situation [25-33]. Students may have access to appropriate physical ideas without using them productively. A concept becomes operative only when the framing of the task and the perceived relevance of available resources allow it to constrain action. Productive laboratory reasoning should therefore be treated as an observable achievement within situated activity, not as a direct consequence of access to information, formulas or AI-generated explanations.

This perspective is closely connected to research on representations and semiotic resources in physics education. Physics knowledge is constructed through the coordinated use of representational, material, computational and discursive resources [25,29]. Since each representation offers only a partial and constrained access to disciplinary meaning, students need support in translating across representations and using them to construct physical interpretation, rather than treating them as self-evident or isolated [34,35]. In laboratory contexts, this coordination is both operational and epistemic: representations, instruments, data outputs and written accounts are used to decide what can be measured, how it should be measured, how results should be interpreted and how claims should be justified.

For this reason, instruction should make decisions and their warrants – the underlying reasons that justify a specific decision or conclusion – visible, rather than present only final procedures, numerical outcomes or polished reports [27,31]. In physics laboratory work, warrants are the reasons that link concepts, representations, measurements or procedures to defensible laboratory decisions and interpretive claims. These warrants form clusters that reveal how students epistemically frame a given situation, and they may shift from moment to moment as students reinterpret the nature of the task and the knowledge relevant to it [30].

These issues become particularly salient in advanced laboratory courses, where tasks are less algorithmic and students must negotiate constraints that resemble authentic disciplinary practice. Experimental competence involves the ability to design procedures, construct and refine models, interpret data, evaluate uncertainty and justify decisions under the constraints of a specific task. This view is consistent with decision-focused accounts of expert problem solving. Price et al. [36] describe science and engineering problem solving as a sequence of decisions in which solvers choose actions by drawing on domain-specific predictive models. Heimbeck et al. [37] extend this perspective to graduate physics coursework, showing that advanced physics tasks require coordinated decisions about representations, approximations, mathematical tools and physical interpretation.

We define *concept-to-decision reasoning* as the process through which learners make a disciplinary concept operative in selecting, justifying or revising an action within a laboratory task. The term synthesises prior work on physics laboratory learning [19-26] and is used here as an analytic construct for describing this specific form of situated laboratory reasoning.

In this sense, a concept is not treated only as a definition, formula or item of declarative knowledge. It becomes educationally significant when it constrains what can reasonably be done, what must be checked, which representation or tool should be trusted and what can count as an adequate measurement or observational claim. In an optical photometry laboratory, for example, airmass becomes operative when it constrains the observing window; seeing when it informs aperture choice; signal-to-noise ratio when it guides exposure settings; ADU and gain when they support reasoning about detector limits and saturation; and comparison and check stars when they validate a differential light curve.

In the present study, concept-to-decision reasoning is not treated as an independent learning outcome or as evidence of learning gain. It is used more narrowly as an analytic specification of epistemic warranting in laboratory activity. A warrant is coded as concept-to-decision reasoning only when a disciplinary concept, representation or measurement constraint is used to justify a concrete observational, analytical or report-level decision. This distinction is important because students may mention correct concepts without making them operative in the laboratory task.

## 2.2 GenAI-mediated physics learning ecology

The distributed character of physics laboratory learning is a clear example of what is called *learning ecology.* In the learning-ecology tradition, learning opportunities are distributed across physical and virtual contexts, activities, tools and social relations [38]. Luckin's ecology of resources extends this view to technology-rich learning environments by focusing on how learners, teachers and artefacts are configured to support mediation [39]. In the present study, this ecological perspective is not used to suggest that every resource has equal educational significance. It is used to indicate that in modern educational contexts, GenAI tools are one resource in a broader configuration that also includes, among the others, the instructor, peers, instruments, databases, software, code, graphs, course materials and assessment.

We define a *GenAI-mediated physics learning ecology* as the situated configuration of human, material, digital and semiotic resources through which students engage in disciplinary learning activity, make decisions, construct artefacts and justify claims, where generative AI functions as one mediating resource among others rather than as a stand-alone tutor. This definition draws on learning ecology and ecology-of-resources perspectives [38,39], on accounts of teacher orchestration in technology-rich learning environments [40], and on sociomaterial analyses that treat educational technologies as part of emergent learning practices rather than neutral channels of information [41,42].

The modifier mediated is essential. In AI-mediated communication, computational agents do not merely transmit messages; they can modify, augment or generate communicative content [43]. In an educational laboratory, this means that GenAI may influence how students formulate questions, organise explanations, interpret evidence and draft reportable claims. The same mediation can support productive inspection of assumptions or encourage premature outsourcing of reasoning. A GenAI-mediated learning ecology must therefore be analysed in terms of both affordances and risks.

This reasoning is analyzed through the lens of AI-mediated epistemic framing. Epistemic framing concerns how students position the AI in relation to knowledge, assumptions, and warrants, while semiotic framing relates to the coordination of representations with physical or measurement problems. Thus, concept-to-decision reasoning emerges at the productive intersection of these two frames, where disciplinary representations become meaningful through justified decision-making.

# 3. Instructional context and AstroTutor design

## 3.1 Structure of the course

The study was conducted in a Master's-level Astrophysics Laboratory course. The optical photometry activity required students to design an observation of a variable star, use the Observation Planner created for the course (see Fig. 2), work with CCD/FITS data, apply aperture and differential photometry, analyse the resulting light curves and prepare a final laboratory report. The reports were submitted as the examination artefact and were assessed within the course, although individual scores and analytic rubrics were not available for this research analysis.

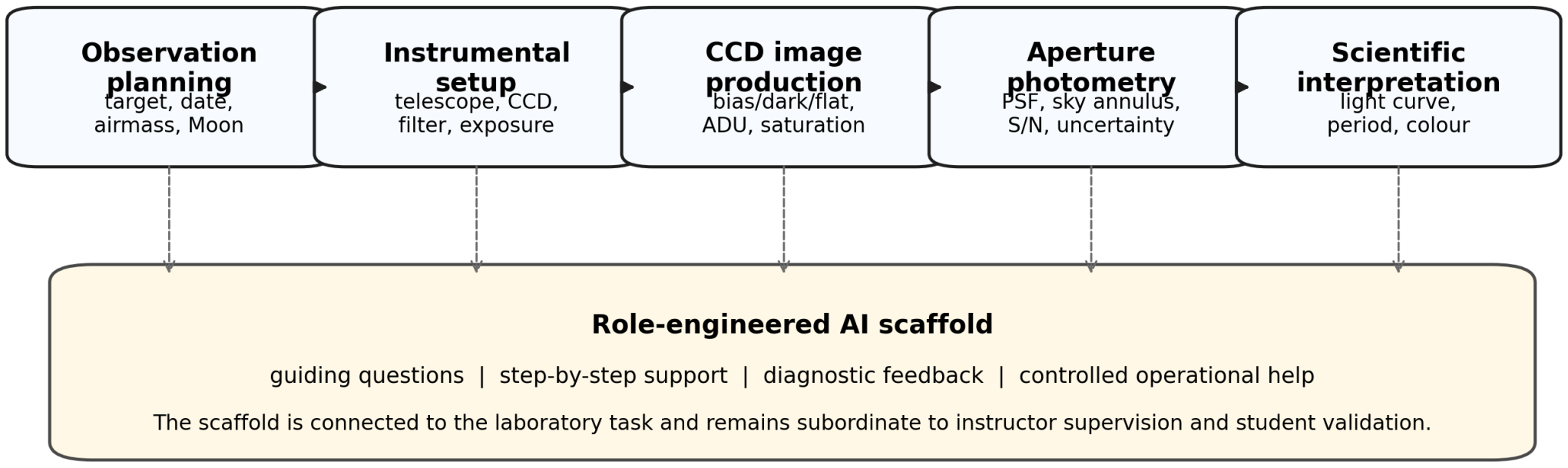

**Fig. 1.** Instructional structure of the optical photometry laboratory. The AI scaffold is designed to support translation across planning, instrumentation, CCD data production, aperture photometry and scientific interpretation while remaining subordinate to student validation and instructor supervision.

The laboratory sequence required students to coordinate several tasks, see Fig. 2. They had to select a suitable variable object, check observability from the local site, estimate whether the expected variation was detectable with the available instrumentation, choose B and V observing strategies when appropriate, avoid saturation, select comparison and check stars, use aperture photometry to extract instrumental magnitudes, construct differential light curves, estimate periods and discuss discrepancies with catalogue or literature values [44-50]. The activity therefore provided a natural context for analysing concept-to-decision reasoning and epistemic warrants.

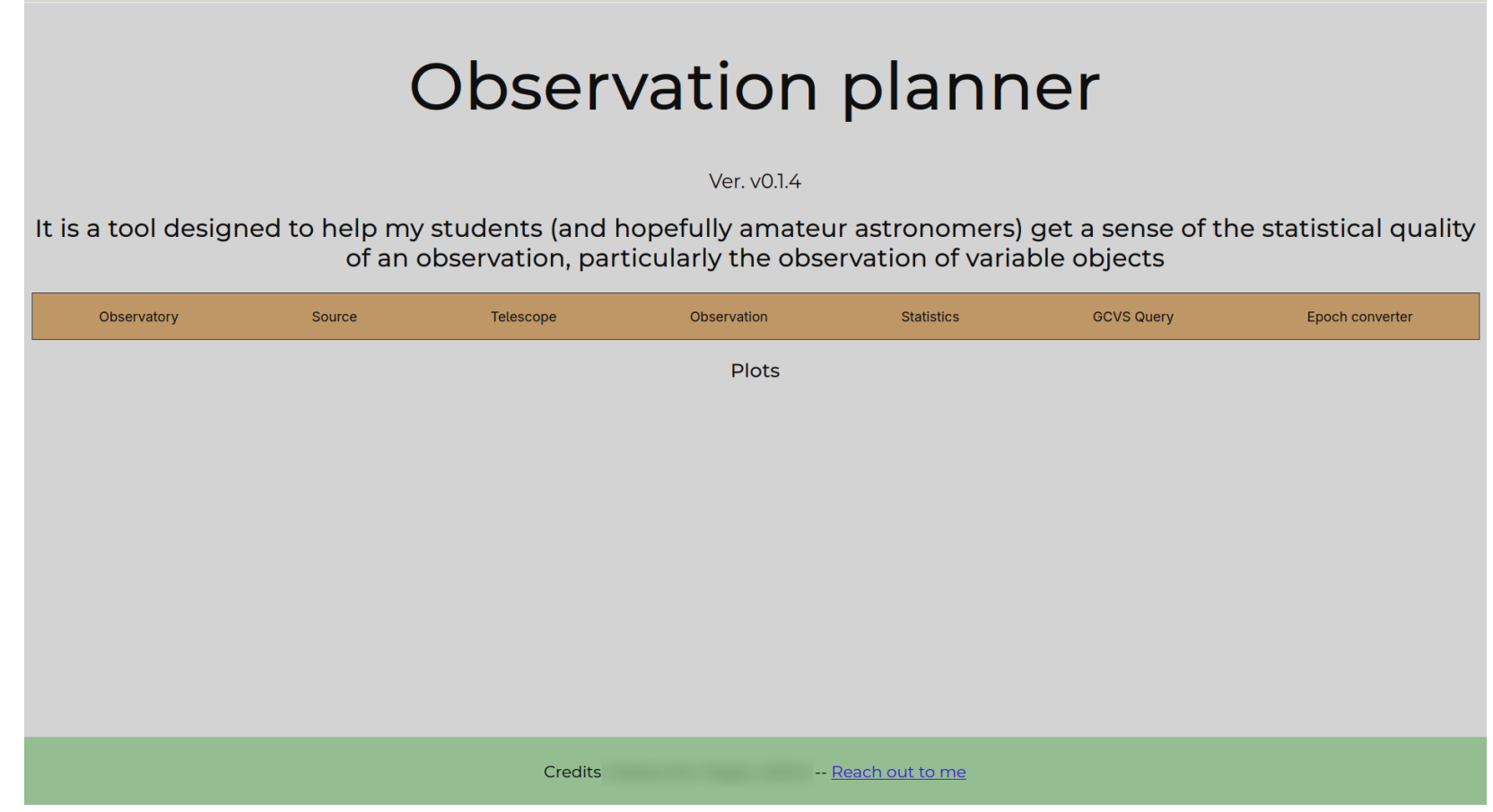

**Fig. 2.** Screenshot of the Observation Planner used in the Optical Photometry module.

Observations were made with the observatory located at the Physics Department of the University of Cagliari. As a tool to support the design of observations and to help students in manage the statistical quality of an observation of variable objects, the lecturer designed an “Observation Planner”, see Fig. 3. It encoded relations among site coordinates, target coordinates, date and time, altitude, azimuth, airmass, sky brightness, seeing, filter, exposure time, sensitivity and saturation. For students, using the planner productively required interpreting interface fields as physical and observational constraints. This made the planner a central semiotic resource in the laboratory ecology.

The final report was used as the assessed artefact for the laboratory activity. Students were required to produce a structured scientific report documenting the full observational and analytical chain developed during the course. The report typically included an introduction to the astrophysical target and its variability class, a theoretical account of the relevant photometric methods, the design of the observation, the criteria used for target visibility and filter selection, the choice of comparison and check stars, the aperture-photometry procedure, the construction of differential light curves, the estimation of the pulsation period, and a discussion of uncertainties, discrepancies with literature values, and possible systematic effects. In this sense, the report was not treated only as a final written product, but as a downstream laboratory artefact in which students were expected to make explicit the conceptual, methodological, representational, and epistemic warrants supporting their observational and analytical decisions.

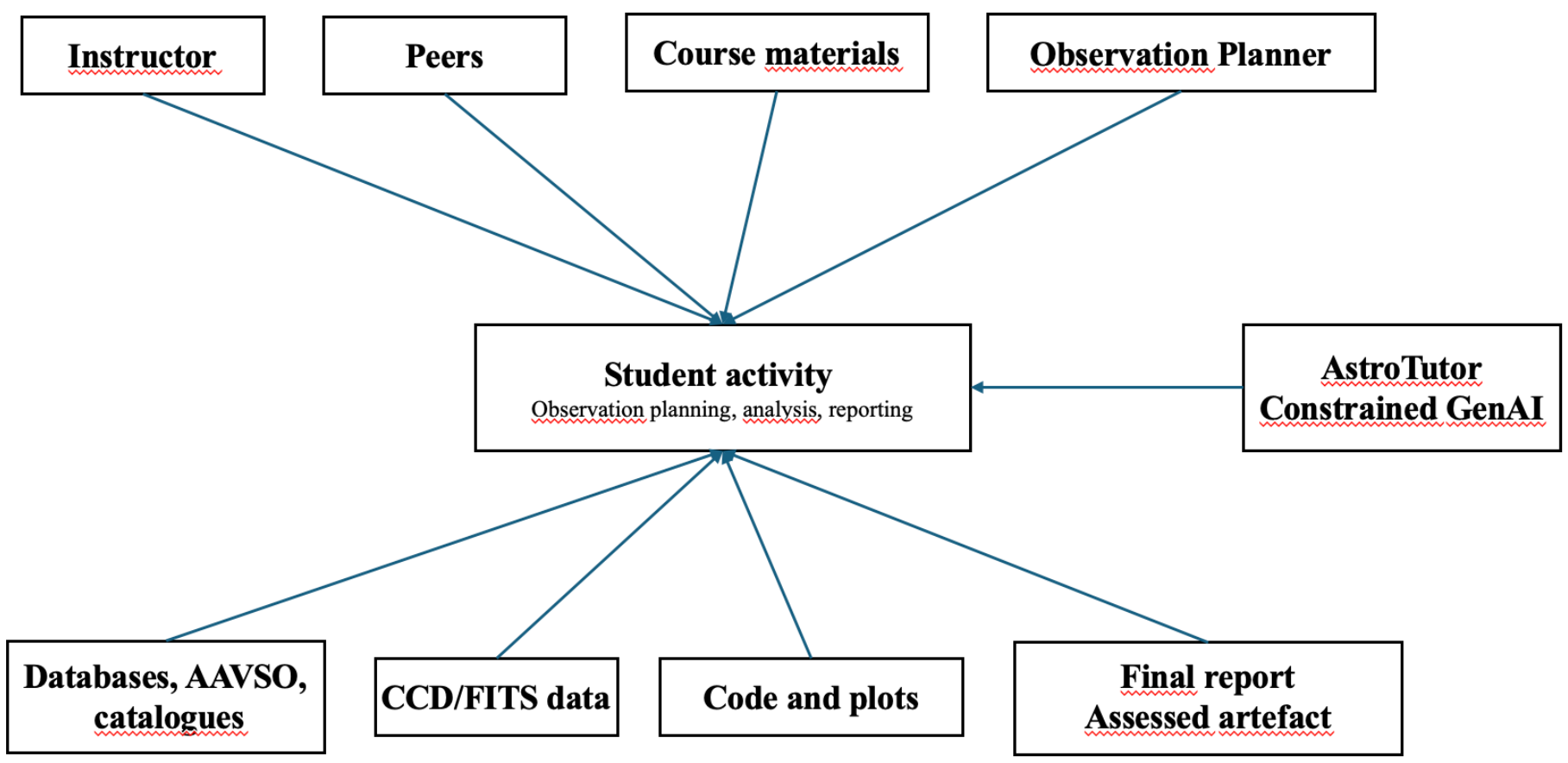

**Fig. 3.** Schematic representation of the GenAI-mediated learning ecology analysed in the study. AstroTutor is treated as one mediating resource within a larger configuration of human, material, digital and semiotic resources.

In the present laboratory, the learning ecology discussed in the previous section had several interacting layers, see Fig. 3. The instructional layer included the lecturer, laboratory sequence, course materials and assessment requirements. The disciplinary-instrumental layer included the telescope context, CCD images, filters, instrumental magnitudes, photometric methods and statistical modelling. The digital-semiotic layer included the Observation Planner, AAVSO and other catalogues, APT or equivalent analysis tools, Python notebooks, plots, formulae and report templates. The social layer included peer collaboration and group authorship. AstroTutor entered this ecology as a dialogic layer: it could translate terms, interpret interface fields, provide formulaic or coding help, review plans and suggest report structures.

### 3.2 AstroTutor

AstroTutor was developed as a custom GPT-based tutor configured around course materials, the optical photometry activity and an instructional script specifying domain, purpose, response modes and limits [51]. It was introduced as an optional support tool. Students were informed that it did not replace the instructor and that they remained responsible for verifying its outputs, interpreting data and authoring their final reports. They interacted with the tool in Italian, but an English translation of the public interface is also available, see Fig. 4.

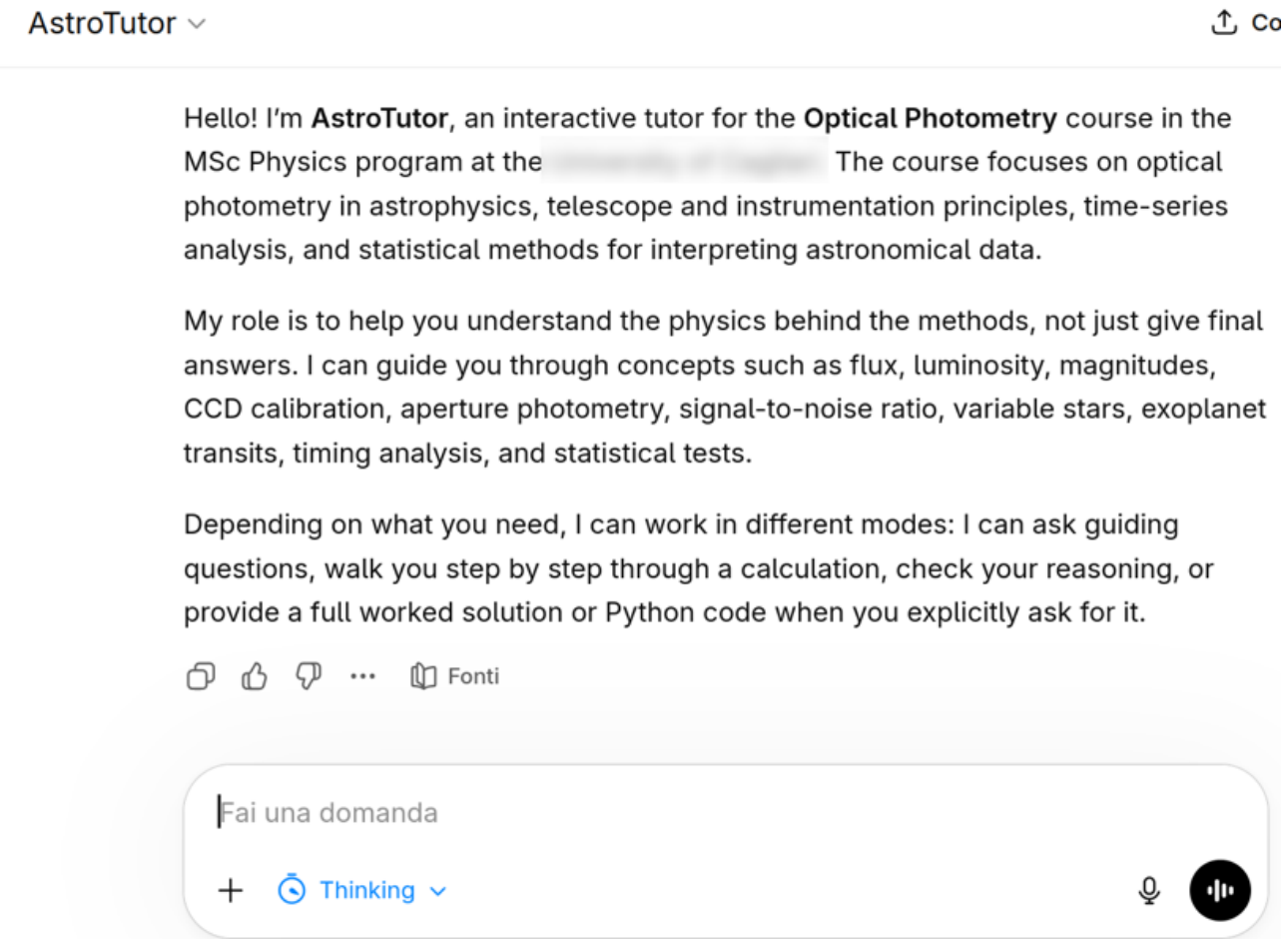

**Fig. 3.** Screenshot of AstroTutor introducing its role as an interactive tutor for the Optical Photometry course. The screenshot illustrates the role-engineered framing presented to students.

The design was intentionally conservative. AstroTutor was intended to support students in understanding the physics behind the laboratory methods, interpreting the Observation Planner, clarifying formulae, reflecting on assumptions, receiving diagnostic feedback on plans or report sections, and obtaining operational help when requested. It was not designed to provide summative evaluation, certify the correctness of work or produce final reports for students.

**Table 1.** AstroTutor design constraints and intended functions in the GenAI-mediated learning ecology.

| Design constraint | Implementation in AstroTutor | Intended function within the learning ecology |
|---|---|---|
| Domain specificity | The tutor was framed around optical photometry, CCD analysis, statistics, time-series interpretation and the course Observation Planner. | Keep dialogue anchored to the astrophysics laboratory rather than to generic AI assistance. |
| Guided reasoning | The tutor could use Socratic or step-by-step explanation when students asked for conceptual clarification. | Support links between concepts and decisions without immediately replacing student reasoning. |
| Interface | The tutor could explain planner fields, hidden variables, | Help students treat software outputs as |

| interpretation | sensitivity curves and observational parameters. | physical representations. |
| --- | --- | --- |
| Diagnostic feedback | Plans, draft sections, plots or code could be discussed in terms of assumptions, omissions and possible refinements. | Make epistemic warrants visible while preserving instructor authority. |
| Controlled operational help | Formulas, calculations and code were allowed when explicitly requested and accompanied by physical interpretation. | Provide practical support without severing calculation from assumptions and verification. |
| Boundary limitation | The tutor was not intended to grade, certify, replace the instructor or complete the report. | Maintain student responsibility, authorship and assessment integrity. |

Finally, ChatGPT was selected as the platform for AstroTutor for both technical and contextual reasons. At the time of design, it allowed the creation of customised GPTs with course-specific materials, role instructions and response constraints [51]. In addition, when explicitly asked during the course, students reported that they were already using ChatGPT mainly to support physics problem solving. AstroTutor was therefore implemented on a familiar platform, but reconfigured as a bounded educational resource.

# 4. Methodology

## 4.1 Research design and corpus

The study adopts an exploratory qualitative case-study design. The aim is not statistical generalisation, causal attribution or measurement of learning gains, but analytical interpretation of how a constrained GenAI tutor was used within an advanced astrophysics laboratory learning ecology and how AI-mediated traces related to downstream artefacts.

AstroTutor was implemented for the first time in the 2025/2026 academic year. Seven students regularly attended the course. The AI-interaction corpus consists of five consenting AstroTutor chat logs from three students. The logs concerned the Observation Planner, coordinate systems, airmass, extinction, seeing, sky brightness, CCD parameters, ADU and gain, S/N, aperture photometry, differential photometry, variable-star observation, formula interpretation, Python support and report writing. The corpus also included a student-produced observing-plan draft pasted into the chat. This artefact provided the strongest evidence of how photometric concepts were assembled into a concrete observing strategy.

The final elaborate consists of the three final group reports submitted as the examination artefact. The reports analysed observations or simulated observations of RV Ari, V0965 Cep and V0536 Peg. Because the reports are group products, they were not assigned to individual students in the analysis.

To complement the chat-log corpus, a brief post-use written reflective questionnaire was administered to the course participants after the course (see Appendix A for details). Data were collected from one week to three months after students gave the exam. The questionnaire was not designed to measure learning gains or to evaluate AstroTutor's effectiveness. Its purpose was to provide limited contextual evidence on aspects of AI use that are only partially visible in chat logs, such as perceived usefulness, response verification, revision of work, over-reliance, and safeguards for future use. The items asked students to describe the laboratory tasks in which AstroTutor was used, episodes of conceptual or methodological clarification, possible connections between photometric concepts and operational decisions, useful or problematic tutor response modes, and strategies for checking AI-generated support against course materials, instructor guidance, the Observation Planner, or observational data. The questionnaire was not analysed as an independent dataset and was not used to support claims about prevalence, learning gains or effectiveness. It was used only as limited contextual evidence about perceived use, non-use, verification and reliability concerns.

**Table 2.** Corpus and analytic status of the available data sources.

| Data source | Available material | Analytic status |
|---|---|---|
| AstroTutor chat logs | Five consenting logs from three students; one off-task conversation excluded from the core analysis. | Primary evidence for AI framing, AI functions, visible warrants and epistemic risks. |
| Final laboratory reports | Three group reports: RV Ari, V0965 Cep and V0536 Peg. | Downstream assessed artefacts; used for traceability analysis, not causal outcome measurement. |
| Reflective questionnaire | Three valid responses, one from an AstroTutor user and one from a non-user. | Limited contextual evidence; not a questionnaire dataset. |
| Course ecology | Observation Planner, course materials, databases, analysis software, code, peers and instructor activity. | Contextual resources; not all directly observed, but necessary for attribution limits. |

Finally, according to the framework of GenAI-mediated ecological learning chat logs can show how students framed AstroTutor and what kinds of reasoning the dialogue made visible. Final reports can show how groups later formulated observation design, data reduction, validation and interpretation. Reflective questionnaires can contextualise perceived use and verification practices. We stress the fact that none of these sources alone provides a controlled measure of individual learning. Their value lies in connecting all the outcomes from tasks and activities to obtain information across the ecology while preserving attribution limits.

## 4.2 Coding and cross-source analysis

The analysis combined qualitative content analysis, thematic analysis and frame analysis [52-54]. The unit of analysis for the logs was the interaction episode, defined as a sequence of turns organised around a specific learning need, see Table 3: interpreting a planner field, clarifying a concept, evaluating a plan, requesting code, checking a formula or asking for report guidance. The unit of analysis for the reports was the report segment, defined as a coherent written unit in which students justified a decision, presented a method, interpreted a plot or evaluated a result.

**Table 3.** Coding frame for AI-mediated functions and downstream artefact analysis.

| Analytic focus | Code | Evidence in chat | Evidence in report | Interpretive caution |
|---|---|---|---|---|
| AI framing | Interface interpreter | Student asks how planner fields, time, coordinates, airmass or sensitivity curves should be understood. | Report uses planner outputs as observing constraints. | Does not prove that interpretation originated in the chat. |
| AI framing | Warrant organiser | Tutor helps connect altitude, S/N, saturation, comparison/check stars or colour to decisions. | Report articulates those connections as methodological justifications. | Specificity of trace determines evidential strength. |
| AI framing | Report scaffold | Tutor suggests structure, missing sections or report-level arguments. | Report contains corresponding architecture or argument sequence. | May reflect support, co-writing or ordinary course expectations. |
| AI risk | Unstable authority | Tutor provides grade-like judgement, overconfident claims or insufficiently grounded feedback. | Report may contain polished claims whose source and validation are unclear. | Requires negative-case analysis rather than positive effectiveness claims. |
| Downstream artefact | Traceable warrant | A specific concept-decision link appears in chat. | The same or transformed warrant appears in the final | Traceability is not causal attribution. |

| | | | report. | |
|---|---|---|---|---|

The coding procedure followed four steps. First, each chat log and each final report were read in full to reconstruct the relevant trajectory of laboratory work. Second, chat episodes were coded according to the role assigned to AstroTutor: conceptual explainer, interface interpreter, operational assistant, diagnostic reviewer, report scaffold, authority to be verified or evaluator. Third, report segments were coded for epistemic warrants: observability/airmass, S/N/exposure/saturation, comparison/check-star validation, aperture/sky-annulus assumptions, fit/model choice, statistical controls, residuals, uncertainty, literature comparison and limitations. Fourth, the chat and report codes were compared to identify traceable, absent, transformed or weakened warrants. For further details in the coding scheme, see Appendix B.

Concept-to-decision reasoning was coded only when a disciplinary concept, representation or measurement constraint was explicitly used to constrain, justify or revise a laboratory decision, methodological choice or interpretive claim. Generic references to concepts were therefore not sufficient for coding. For example, seeing was coded only when it was connected to PSF width, aperture radius, sky contribution, signal-to-noise ratio or exposure choices. Similarly, S/N was coded only when it constrained exposure time, source magnitude limits, detector noise, saturation, uncertainty or comparison among stars. References to comparison and check stars were coded only when they were linked to the validity of differential photometry, for instance through stability, angular proximity, colour index compatibility or comparison-check consistency.

For each candidate chat episode or report segment, the analysis followed four guiding questions: which concept or representation was invoked; which laboratory decision, methodological choice or interpretive claim it constrained; what warrant connected the concept to that decision; and whether the same warrant was visible, transformed, weakened or absent in another artefact. This procedure was intended to distinguish concept-to-decision reasoning from general conceptual talk. Thus, a definition of airmass was not coded as concept-to-decision reasoning unless it was connected to an observing window, elevation threshold or extinction-related limitation. Likewise, a definition of S/N was not sufficient unless it was used to justify exposure time, detectability, uncertainty control, saturation avoidance or the comparability of target, comparison and check stars.

### 4.3 Laboratory artefacts

The final reports are analysed as downstream laboratory artefacts. They aim was to investigate the emergence of specific epistemic warrants and in which way they can influence representations or decisions in AI-mediated interactions. A relation was coded as strong only when a specific artefact or decision appeared in both the chat with AstroTutor and the report. Generic similarity between a tutor explanation and a report section was not interpreted as evidence of AI uptake.

The analysis therefore focuses on warrants that are central to the laboratory activity. For example, the warrant for selecting an observing window may involve altitude, airmass, and extinction; the warrant for selecting comparison and check stars may involve angular proximity, colour similarity, magnitude range, non-variability, and comparison-check stability; and the warrant for interpreting a period estimate may involve fit quality, residuals, uncertainty, p-values, agreement across filters, and comparison with catalogue values. These warrants provide an analytical lens for examining how students' laboratory reasoning is documented across AI-mediated interactions and final assessed reports, without attributing such reasoning causally to AstroTutor.

### 4.4 Trustworthiness and ethics

Trustworthiness was addressed through explicit unit definitions, cross-source comparison, independent coding, analytic consensus, conservative interpretation and negative-case analysis [52-54]. The unit of analysis for chat logs was the interaction episode, defined as a sequence of turns organised around a specific learning need. The unit of analysis for final reports was the report segment, defined as a coherent written unit in which students justified a decision, presented a method, interpreted a plot or evaluated a result. These definitions were used to prevent generic thematic similarity from being treated as evidence of AI uptake.

Both authors first read the full corpus independently and coded the chat logs and report segments using the coding frame described in Section 4.2. The first round of coding focused on the role assigned to AstroTutor in each episode, including conceptual explainer, interface interpreter, operational assistant, diagnostic reviewer, report scaffold and authority to be verified. The second round focused on epistemic warrants in the reports, including observability and airmass, S/N and exposure, saturation, comparison/check-star validation, aperture and sky-annulus assumptions, fit/model choice, residuals, uncertainty, literature comparison and limitations.

After the independent coding, the authors compared the assigned codes episode by episode and segment by segment. Divergences were resolved through discussion rather than by numerical aggregation, because the aim was interpretive consistency rather than statistical reliability. Disagreements were used to refine code definitions, distinguish strong from weak traceability, and identify cases in which a report-level warrant could plausibly derive from ordinary course design, peer collaboration or instructor guidance rather than from AstroTutor. A relation was therefore coded as strong only when a specific artefact, decision or warrant appeared in both the chat and the final report. Generic overlap between a tutor explanation and a report section was not interpreted as evidence of AI influence.

This procedure was intended to make the interpretive chain inspectable and to distinguish credibility, transparency and transferability from claims of causal effectiveness [55-57]. The analysis therefore treats the available corpus as situated evidence of AI-mediated activity, not as a basis for prevalence estimates, outcome measurement or individual performance evaluation.

## 5. Results and discussion

AstroTutor entered the laboratory ecology through several task-dependent framings: interface interpreter, warrant organiser, report scaffold, conceptual and Socratic rehearsal partner, operational assistant, unstable authority and resource whose traces may appear in downstream reports. See also Table 3 and 4 for sake of completeness.

**Table 4.** Cross-source synthesis of AI-mediated frames and downstream report traces.

| Source | Trace evidence | Interpretive status |
|---|---|---|
| RV Ari | Report includes elevation > 30 degrees, comparison/check stars, S/N reasoning and systematic-discrepancy discussion. | Weak to moderate; mainly standard course reasoning and group authorship. |
| V0965 Cep | Chat and report share planner interpretation, airmass/elevation, sensitivity, saturation and validation concerns. | Moderate; report includes analyses not visible in the chats. |
| V0536 Peg | A pasted observing plan is recognisable in the final report architecture, B/V strategy, planner rationale and comparison/check validation. | Strongest trace; plan-report continuity, not causal evidence of learning. |
| Reflective responses | Two users report different framings: verification-oriented support with hallucinations; Socratic conceptual "ping-pong" without direct experimental use. One | Contextual only; clarifies perceived use, risk and adoption boundaries. |

| | non-user preferred another platform. | |
|---|---|---|

## 5.1 AI as interface interpreter: the planner becomes a conceptual object

The clearest interface-interpreter frame occurred when a student opened a conversation by explaining that the immediate goal was to understand the Observation Planner, explore the site, identify what each section did and write notes connected to theoretical concepts from the course. The student wrote:

*"My objective is to orient myself in the Observation Planner, created by the professor to select observable celestial objects and correctly design the observation. I want to explore the site, understand what each part does, and write notes that recall the theoretical concepts seen in class."*

This prompt is analytically important because the student did not frame AstroTutor as a generic answer source. The student positioned the software interface itself as something to be interpreted physically. The semiotic object was the planner: fields, tabs, plots and numerical constraints. The epistemic task was to understand what those interface elements meant for selecting an observable target.

The same episode developed when the student identified weather, target-observer geometry, brightness, exposure time and saturation as factors affecting observability. AstroTutor reorganised these into atmospheric, geometrical and instrumental factors and asked what happens when a star approaches the horizon. The student answered:

*"If a star is close to the horizon, it has to cross a greater atmospheric thickness; the photons have had more opportunity to interact with atmospheric particles."*

This is not evidence of learning by itself, but it is evidence of a productive framing. The student moved from a planner question to a physical account of why altitude matters. The tutor continued by connecting this intuition to airmass, the plane-parallel approximation $X \simeq \sec z$, extinction, colour effects and PSF degradation. The important result is not that the tutor supplied a correct definition of airmass. The important result is that the interaction made explicit how a planner quantity can become an observing constraint.

The new reflective response adds a related but more limited case. The student reported using AstroTutor during observation planning to revise basic concepts seen in class, including airmass, photometry, signal-to-noise ratio and filters. However, the same student explicitly stated: *"I did not use AstroTutor for the experiment."* This distinction is important. AstroTutor could support the interpretation and consolidation of conceptual resources, but that does not automatically imply that those resources were used to make operational decisions in the laboratory.

Downstream reports show that planner outputs became part of report-level observation design. The V0965 Cep report states that the source was checked with the Observation Planner, that the exposure did not saturate across the observing window, and that the airmass remained $X \leq 2$, corresponding to an elevation above 30 degrees. The V0536 Peg report similarly uses the Observation Planner to impose an elevation threshold above 30 degrees and to verify that the magnitude variation was well above three times the sensitivity curve. The RV Ari report also reports an elevation threshold above 30 degrees verified with the course planner.

This is a meaningful relation, but it must be stated carefully. The reports show that planner quantities were translated into observing constraints. They do not show that AstroTutor caused this translation. The relation is strongest for the student who explicitly used AstroTutor to interpret planner fields and for the V0536 Peg plan-report continuity; it is weaker where report criteria may derive from course design, peer work or ordinary laboratory instruction.

## 5.2 AI as warrant organiser: from concepts to defensible decisions

A second function was warrant organisation. AstroTutor was useful when it helped connect a concept, representation or parameter to the reason for a decision. In the corpus, the most recurrent warrants concerned observability, exposure, saturation, S/N, comparison/check-star selection and validation.

To make this distinction explicit, Table 5 summarises the main concept-to-decision warrants identified across the corpus. The table does not report frequencies and is not intended to attribute learning outcomes to AstroTutor. Rather, it shows how specific disciplinary concepts or representations became analytically relevant only when they were connected to laboratory decisions, methodological choices or interpretive claims. In this sense, the table clarifies how concept-to-decision reasoning was used to distinguish strong epistemic warrants from generic conceptual references.

**Table 5.** The table shows concept-to-decision warrants across AI-mediated interactions and final reports. It indicates where disciplinary concepts were connected to laboratory decisions, while specifying trace status without implying causal attribution to AstroTutor.

| Concept or representation | Decision constrained | Warrant type | Trace status |
|---|---|---|---|
| Airmass / elevation | Observing window and elevation threshold | Observability and atmospheric path length | Visible in chats and reports; strongest when linked to planner interpretation |
| Sensitivity curve / S/N | Detectability and exposure strategy | Signal relative to noise and expected magnitude variation | Moderate to strong in V0536 Peg and V0965 Cep |
| ADU / gain / saturation | Maximum exposure and detector linearity | Detector response limits | Visible in report-level justification; partial AI trace |
| Seeing / PSF / sky background | Aperture choice and photometric extraction | Relation between image quality, aperture radius, sky contribution and noise | Present in AI feedback; weaker downstream trace |
| Comparison and check stars | Validity of differential photometry | Stability, proximity, colour compatibility and control of atmospheric/systematic effects | Strong in final reports; strongest where plan-report continuity is visible |
| Fit quality / residuals / literature values | Interpretation of period and discrepancies | Model adequacy and uncertainty control | Mainly report-level; weak AI trace |

The V0536 Peg plan submitted in the chat already contained a substantial observing rationale: B and V light curves, period, colour, target coordinates, AAVSO information, visibility dates, elevation threshold, lunar and sky-brightness considerations, exposure estimates, cadence, comparison and check stars, and a reduction strategy. AstroTutor's diagnostic feedback did not create this plan from nothing. It identified missing or weak warrants: the need to state assumptions more explicitly, link seeing to aperture choice, verify comparison-star stability a posteriori, justify colour similarity and differential extinction, and treat sky background, read noise and dark current as limits of simplified exposure calculations.

The final V0536 Peg report contains several of these warrant types. It uses the Observation Planner and sensitivity curve to justify observability; it discusses exposure time and saturation through CCD linearity; it introduces differential photometry through the cancellation of atmospheric effects under proximity assumptions; it validates the reference star through a check star; and it moves from harmonic fits to a global simultaneous model when an unexpected long-term trend appears.

The V0965 Cep report provides a moderate parallel case. It includes an Observation Planner sensitivity curve, a statement that the selected source remained within the saturation and airmass constraints, a comparison-star selection process using AAVSO data, checks of colour compatibility, and statistical validation of comparison/check stability. These report-level warrants overlap with the type of reasoning made visible in the planner and operational-support chats, but the report also contains advanced analyses that cannot be traced to a specific AI exchange.

The RV Ari report shows similar concept-to-decision reasoning: selection of the target based on period, magnitude, variation amplitude and elevation; selection of comparison and check stars based on angular proximity, magnitude, colour and stability; discussion of S/N comparability across target, comparison and check stars; and recognition that discrepancies with literature may indicate systematic error. However, the available AI trace is too weak to attribute these warrants to AstroTutor.

The reflective responses qualify the warrant-organiser interpretation in two ways. One user explicitly denied that AstroTutor helped connect concepts to operational decisions, stating that it "*only provided information.*" The new user reported a strong sense of active conceptual learning through Socratic dialogue, but also stated that AstroTutor was not used for the experiment. Thus, the questionnaire evidence does not strengthen the claim that AstroTutor supported operational decision-making. It instead shows that students may experience the tutor as useful for conceptual rehearsal even when no direct concept-to-decision trace is visible in the assessed artefacts.

These cases clarify the role of concept-to-decision reasoning in the analysis. The relevant evidence is not the mere presence of correct photometric terminology, but the use of that terminology to constrain an observing, analytical or reporting decision. In this sense, concept-to-decision reasoning functions as a criterion for distinguishing stronger epistemic warrants from weaker conceptual references. The strongest cases occurred when students brought concrete artefacts into the AI dialogue, such as the Observation Planner or an observing plan. The weakest cases occurred when AstroTutor was used mainly for definitions, conceptual rehearsal or generic explanation without a visible connection to operational decisions or downstream report claims.

### 5.3 AI as report scaffold: useful support and authorship risk

AstroTutor also functioned as a report scaffold. Students asked for help evaluating plans, organising report sections, connecting code to analysis and deciding what should be included in the written account. This function is pedagogically plausible in an advanced laboratory, where the report is not merely a record of results but a scientific artefact requiring argument structure, methods, uncertainty, plots and discussion of limitations.

The V0536 Peg case shows the strongest plan-report continuity. In the chat, the student pasted a complete LaTeX observing plan and asked AstroTutor whether the observation was adequate and what should be added. The tutor identified strengths and suggested refinements. The final report later retains the same object, B/V strategy, observation rationale, planner-based sensitivity argument, elevation threshold, comparison/check framework and model-based period analysis. This relation suggests that AI-mediated diagnostic feedback may participate in the transformation of an observing plan into a more articulated report architecture.

The reflective evidence supports a boundary-sensitive interpretation of this function. One user stated that AstroTutor should be used for checking, clarifying and improving material already produced by the student, not for replacing the student's work, providing complete reports or simulating grades. The new user similarly argued that AstroTutor should not replace lectures or interaction with the instructor and should mainly be used in Socratic mode on topics already treated in class. These responses align with the intended design of AstroTutor, but they also confirm that report scaffolding is an assessment-sensitive function.

This function therefore creates the most delicate interpretive problem. Report scaffolding can support learning if students use it to inspect and revise their own reasoning. It can also weaken authorship if students treat the generated structure as a substitute for epistemic work. The available data do not allow us to distinguish fully among these possibilities. The analysis therefore treats report scaffolding as an AI function and an assessment-boundary risk, not as evidence of improved learning.

The reports also show that AI support was not limited to AstroTutor. In the V0536 Peg report, the figure illustrating aperture-photometry regions is explicitly described as generated with Google Gemini. This is important for the ecological interpretation. The laboratory did not contain a single AI intervention. Students could draw on multiple AI and non-AI resources while constructing the final artefact. Any claim about AstroTutor must therefore be restricted to traceable interactions, not to the entire report.

### 5.4 AI as unstable authority: verification burden and role drift

The fourth function is negative but central. AstroTutor sometimes moved beyond scaffolded support toward authority. The clearest case occurred when the tutor, after reviewing an observing plan, suggested a possible laboratory mark in the range 28-30/30. This was inconsistent with the intended role of the system. The tutor was designed to provide formative diagnostic feedback, not assessment certification. The grade-like response is therefore coded as role drift.

This episode matters because it exposes a boundary problem that is easy to underestimate. In an assessed laboratory activity, students may ask AI systems whether a plan, code or report is "good enough". A fluent response can appear as evaluative authority even if it lacks access to instructor criteria, grading standards, course-specific expectations or the full experimental context. This risk is not only technical. It is epistemic and institutional: the system may blur the distinction between formative feedback, self-assessment, instructor judgement and summative evaluation.

The reflective responses reinforce this caution. One user reported "too many hallucinations" when asking for definitions of important physical concepts and therefore treated verification against external sources as necessary. The new user reported a more specific disciplinary problem: AstroTutor formulated a magnitude difference as a difference between two logarithms, producing logarithmic arguments that were not dimensionless, whereas the instructor had emphasised that the expression should remain a single logarithm of a ratio. This is a valuable negative case because it concerns not only factual accuracy, but also the mathematical and dimensional form of a physics representation.

These responses cannot be generalised, but they identify the verification burden created by GenAI support in a physics laboratory. The same tool that can organise dialogue and support conceptual rehearsal can also produce plausible but problematic formulations. Therefore, AstroTutor should be interpreted as a resource requiring disciplinary control, not as an autonomous authority.

### 5.5 Reports as downstream artefacts: traceability without causal attribution

The expanded corpus does not support a statistical correlation between AstroTutor use and learning as measured by the final report. There are only three reports, the reports are group products, group membership mixes users and non-users, and no individual grade or rubric scores were available.

The reports do, however, support a downstream traceability analysis. In the V0536 Peg case, the trace relation is strongest because a concrete observing plan was pasted into the tutor and the final report later contains a recognisable continuation of the same plan architecture. In the V0965 Cep case, the relation is moderate: the chat and report share attention to planner interpretation, airmass/elevation thresholds, sensitivity curves, S/N, saturation and differential-photometry validation, but the report develops analyses that go beyond the available chat. In the RV Ari case, the relation is weak to moderate: the report contains strong photometric reasoning, but the available AI trace does not distinguish AstroTutor influence from course instruction and group work.

Several warrants appear across reports regardless of the strength of AI trace: elevation above 30 degrees, sensitivity or detectability, saturation avoidance, comparison/check-star selection, S/N comparability, aperture-photometry assumptions, statistical validation and discussion of discrepancies with literature values. These are educationally important because they show that the reports did not merely reproduce definitions.

They articulated relations between concepts and decisions. Yet their presence across reports also weakens any simple claim about AI-specific effects, because such warrants are part of the disciplinary task design.

Reflective response in the questionnaire strengthens this methodological caution rather than the causal claim. A student described AstroTutor as useful for active conceptual rehearsal, but also stated that its use did not extend to the experiment and only indirectly related to the final report through a colleague's use. This confirms that perceived usefulness, AI-mediated interaction and report-level traces are analytically distinct.

The most defensible result is therefore conditional: AI-mediated dialogue was most informative when students used it around concrete artefacts and local decisions. Interface interpretation and diagnostic review produced the clearest traceable relations to reports. General conceptual explanation, Socratic rehearsal and operational help produced weaker trace evidence unless linked to assumptions, representations and verification.

### 5.6 Further AI-mediated functions in the laboratory ecology

Across the corpus, AstroTutor did not function as a single pedagogical object. Its role changed according to how students framed the interaction and to the local task within the laboratory ecology. Four recurrent functions can be distinguished beyond the strongest interface and report-trace cases.

First, AstroTutor acted as a conceptual explainer when students asked short definitional questions about photometry, optical photometry, aperture photometry, differential photometry, check stars, airmass, S/N and filters. These exchanges organised terminology into a coherent measurement hierarchy, but they provide weak evidence of uptake unless definitions were later used to justify a concrete observational or analytical decision.

Second, AstroTutor acted as a Socratic rehearsal partner. The new reflective response is the clearest evidence for this frame: the student described a "ping-pong dynamic" in which they reformulated concepts from the course, received feedback, and felt that they were actively approaching a more complete view of the topic. This is a relevant learning-ecology function, but it remains distinct from documented laboratory decision-making.

Third, AstroTutor acted as an operational assistant and semiotic translator when students asked about formulas, code, ADU, gain, S/N, seeing and exposure estimates. This function was potentially productive when calculations were linked to assumptions and verification, for example when seeing was related to PSF width and aperture choice, or when sky brightness was interpreted as a noise contribution.

Fourth, the tutor acted as a diagnostic reviewer when a student submitted a complete observing plan for V0536 Peg, including scientific aims, visibility constraints, B and V magnitudes, exposure estimates, a B-V sequence, comparison/check stars and a data-reduction strategy. In this case, AstroTutor made several epistemic controls explicit, including sky-brightness assumptions, lower-limit exposure estimates, comparison-star stability and differential-extinction limits.

These cases show that productive AI-mediated interaction was conditional rather than automatic. It was most visible when the tutor helped students connect a semiotic resource to a physical concept and then to an operational or epistemic decision: planner fields to observability and airmass; seeing and sky brightness to aperture choice and S/N; comparison/check stars to the credibility of differential photometry; cadence and filters to the reconstruction of a light curve. Conversely, when the tutor was framed as a generic explainer, shortcut or evaluator, the interaction either provided only weak evidence of concept-to-decision reasoning or introduced assessment-boundary risks. Within the GenAI-mediated laboratory ecology, AstroTutor is therefore best described as a contingent epistemic resource: it can support conceptual rehearsal and warrant organisation, but it can also shift verification responsibility back to students and instructors.

## 6. Educational implications

Our findings suggest that the GenAI-mediated learning ecology framework can contribute to physics education research by shifting the focus from the general effectiveness of an AI tool to its role in disciplinary mediation. This shift is important because physics laboratory learning is not reducible to receiving correct explanations, but involves designing procedures, coordinating instruments and representations, evaluating uncertainty, interpreting data and justifying claims [19-26]. In this perspective, research should focus on the role that GenAI tutors can have in the ecology of laboratory reasoning: which resources they connect, which warrants they make inspectable, which responsibilities they redistribute, and which assessment boundaries they threaten.

For advanced physics laboratories, the main implication is that GenAI tutors should be designed around disciplinary artefacts and epistemic decisions, not around generic explanation alone. In the present corpus, the most informative interactions occurred when AstroTutor was connected to the Observation Planner, observing plans, sensitivity estimates, comparison/check-star choices, exposure decisions, code, plots or report drafts. This is consistent with research showing that advanced laboratory work requires students to make and justify decisions under constraints, rather than simply apply known formulas [23-26]. It is also consistent with framing research: a concept becomes educationally productive when students treat it as relevant to the activity and use it as a warrant for action [27,28,31-33].

As suggested by the reflective responses in the questionnaire, AstroTutor should also support Socratic conceptual rehearsal, but this mode should be kept distinct from experimental decision-making and report production [15]. A useful design principle is that the tutor should first ask students to reformulate a concept, identify the decision at stake, and state the evidence or assumption they are using. Direct explanations and step-by-step solutions may be appropriate when students are dealing with unfamiliar or difficult material, but they should be accompanied by prompts that return responsibility to the student: What have you checked? Which representation supports the claim? What assumption would make this conclusion invalid?

This also clarifies the role of the instructor. In technology-rich learning environments, teaching involves orchestrating tools, tasks, representations and interactions rather than merely providing access to digital resources [40]. In a GenAI-mediated astrophysics laboratory, this orchestration includes defining legitimate AI functions, requiring verification practices, specifying how AI use should be declared, and designing assessment criteria that reward epistemic control rather than polished prose. The instructor's role is therefore not displaced by AI; it becomes more explicitly focused on designing epistemic constraints, validation routines and assessment boundaries.

A second implication concerns prompting. Students should not only be taught how to ask for definitions, but how to formulate warrant-oriented prompts. In an advanced laboratory, "What is airmass?" is less educationally specific than "How does airmass constrain my observing window, and what elevation threshold is defensible for this target?" Similarly, "What is S/N?" is less productive than "How do source magnitude, exposure time, sky background, read noise and saturation constrain my measurement strategy?" Such prompts push the AI interaction toward concept-to-decision reasoning, where physical quantities become constraints on observational or analytical choices.

A third implication concerns verification. The reflective responses show that errors may concern both definitions and representational form, as in the reported case of magnitude notation. This suggests that AI tutors in physics laboratories should include explicit prompts for dimensional consistency, formula interpretation, assumptions, comparison with course materials and validation against data or instructor criteria. Reliability cannot be treated as a generic warning; it must be translated into disciplinary checking routines.

A fourth implication concerns feedback and report writing. AI-based diagnostic feedback may be valuable when it asks students to clarify assumptions, compare alternatives, check consistency, or justify methodological choices. This is compatible with research on feedback and self-regulated learning, where productive feedback supports students in closing the gap between current and desired performance without replacing their responsibility for judgement [58,59]. However, in assessed laboratory activities, AI-supported report production creates a specific risk: coherent prose may conceal whether students have reconstructed the reasoning, verified assumptions, or merely accepted a generated structure. Reports should therefore require explicit reconstruction of epistemic warrants: why a target was observable, why a comparison star was acceptable, why an exposure time was defensible, why a fit was adequate, and what limitations remain.

Finally, AI systems used in laboratory courses should not provide grades, grade ranges, certification language or final judgements that simulate the instructor's evaluative authority. This is not a minor design preference but a condition for maintaining the distinction between formative support and summative assessment. The safest role for AstroTutor is therefore diagnostic and dialogic: it should help students inspect their own reasoning, identify missing checks and improve already produced material, while leaving understanding, authorship, final argumentation and evaluation under the responsibility of students and instructors.

## 7. Conclusion

This study examined AstroTutor as one mediating resource within a GenAI-mediated learning ecology in an advanced astrophysics laboratory. The analysis examined how students used AstroTutor alongside course materials, the Observation Planner, astronomical databases, software tools, code, peer collaboration, instructor expectations and the final assessed report. This ecological perspective was necessary because the educational meaning of the tutor changed according to the task, the artefact being produced and the other resources involved.

In response to RQ1, AstroTutor was framed in multiple ways. These were not fixed student profiles but task-dependent framings. The tutor acted as an interface interpreter when students worked with the Observation Planner, as a diagnostic reviewer and report scaffold when they submitted observing plans, as a conceptual explainer for definitions, as an operational assistant for code or calculations, and as a Socratic rehearsal partner when students used it to reformulate and check concepts already treated in class. The new reflective response is important because it identifies this Socratic use as subjectively valuable while also separating it from direct use in the experiment.

In response to RQ2, several epistemic warrants were traceable across chats and reports, especially when students brought concrete artefacts, plans or representations into the AI dialogue. Some warrants were transformed across artefacts; others were absent or weakened. The questionnaire responses help in clarifying that perceived usefulness for conceptual understanding may coexist with no direct trace in experimental decision-making or report writing. The reports should therefore be read as products of the broader laboratory ecology, not as outcomes attributable to AstroTutor alone.

These findings indicate that GenAI use in advanced physics laboratories must be explicitly designed, bounded and visible. Students need guidance on appropriate uses—conceptual clarification, Socratic rehearsal, interface interpretation, code debugging and diagnostic questioning—and on prohibited uses, such as replacing analysis, certifying correctness, simulating assessment or producing final reports. Instructors should design prompts, rubrics and report requirements that make epistemic responsibility visible.

The study has significant limitations. The corpus is small and voluntary; only three students provided analysable AstroTutor logs; the reports are group products; and no individual grades, rubric scores, pre/post measures, classroom observations, screen recordings or report revision histories are available. The questionnaire contains only three valid responses, two from AstroTutor users and one from a non-user, and cannot be treated as a survey dataset. These constraints prevent claims about prevalence, effectiveness, learning gains or causal impact.

The analysis is also limited by the partial nature of chat logs. A log records what students typed and what the tutor produced, but not silent reasoning, peer discussion, instructor feedback, drafting history or decisions made outside the chat. A student may use an AI response critically, ignore it, transform it or copy it; without interviews and revision histories, these possibilities cannot be distinguished reliably. Future studies should examine the GenAI-mediated learning ecology more systematically, investigating how different AI framings—Socratic rehearsal, interface interpretation, diagnostic review and operational support—may support or constrain students' physics learning in advanced laboratory courses.

## Acknowledgments

The authors acknowledge all the students attending the Laboratorio di Astrofisica (Astrophysics Laboratory) course for the Master's-level physics curriculum at the University of Cagliari (UniCA).

## Data availability statement

The anonymised excerpts reported in the paper are derived from student-AI chat logs and student-produced reports collected with informed consent or as part of the course documentation. The full logs and reports contain course-specific and potentially identifying contextual information and are therefore not publicly released. Additional anonymised excerpts may be made available by the corresponding author upon reasonable request, subject to consent and local data-protection requirements.

## Ethical statement

Informed consent to participate in the study has been obtained from participants. Any identifiable individuals participating at the study have been also aware of intended publication. Informed consent to publish has been obtained from participants of the study. This work was carried out in accordance with the principles outlined in the journal's ethical policy and with the "Codice etico e di comportamento" of the University of Cagliari.

## Declaration of generative AI in scientific writing

The authors declare that no generative artificial intelligence tools were used in the design, analysis, or interpretation of this study. Language editing assistance was limited to grammar and style correction using AI-assisted tools (ChatGPT-5), with all intellectual content, analysis, and interpretation solely the responsibility of the authors.

## Conflict of interest

There are no known conflicts of interest associated with this publication, and there has been no significant financial support for this work that could have influenced its outcome.

## Orcid ID

M Tuveri https://orcid.org/0000-0001-5686-1713
A Riggio https://orcid.org/0000-0002-6145-9224

## References

[1] Holmes, W., & Tuomi. (2022). I. State of the art and practice in AI in education. European Journal of Education 57, 542-570 https://doi.org/10.1111/ejed.12533

[2] Zawacki-Richter, O., Marin, V. I., Bond, M., & Gouverneur, F. (2019). Systematic review of research on artificial intelligence applications in higher education: Where are the educators? International Journal of Educational Technology in Higher Education 16, 39 https://doi.org/10.1186/s41239-019-0171-0

[3] Kasneci, E. Sessler, K., Küchemann, S., Bannert, M., Dementieva, D., Fischer, F., Gasser, U., Groh, G., Günnemann, S., Hüllermeier, E., Krusche, S., Kutyniok, G., Michaeli, T., Nerdel, C., Pfeffer, J., Poquet, O., Sailer, M., Schmidt, A., Seidel, T., Stadler, M., Weller, J., Kuhn, J., & Kasneci, G. (2023). ChatGPT for good? On opportunities and challenges of large language models for education. Learning and Individual Differences 103, 102274 https://doi.org/10.1016/j.lindif.2023.102274

[4] Shi, Y., Yu, K., Dong, Y., & Chen, F. (2025). Large language models in education: A systematic review of empirical applications, benefits, and challenges. Computers and Education: Artificial Intelligence 10, 100529 https://doi.org/10.1016/j.caeai.2025.100529

[5] Dahlkemper, M. N., Lahme, S. Z., & Klein, P. (2023). How do physics students evaluate artificial intelligence responses on comprehension questions? A study on the perceived scientific accuracy and linguistic quality of ChatGPT. Physical Review Physics Education Research 19, 010142 https://doi.org/10.1103/PhysRevPhysEducRes.19.010142

[6] Polverini, G., & Gregorcic, B. (2024). How understanding large language models can inform the use of ChatGPT in physics education. European Journal of Physics 45, 025701 https://doi.org/10.1088/1361-6404/ad1420

[7] Kortemeyer, G. (2023). Could an artificial-intelligence agent pass an introductory physics course? Physical Review Physics Education Research 19, 010132 https://doi.org/10.1103/PhysRevPhysEducRes.19.010132

[8] Lo, C. K., Hew, K. F., & Jong, M. S. Y. (2024). The influence of ChatGPT on student engagement: A systematic review and future research agenda. Computers & Education 219, 105100 https://doi.org/10.1016/j.compedu.2024.105100

[9] VanLehn, K. (2011). The relative effectiveness of human tutoring, intelligent tutoring systems, and other tutoring systems. Educational Psychologist 46, 197-221 https://doi.org/10.1080/00461520.2011.611369

[10] Graesser, A. C., Chipman, P., Haynes, B. C., & Olney, (2005). A. AutoTutor: An intelligent tutoring system with mixed-initiative dialogue. IEEE Transactions on Education 48, 612-618 https://doi.org/10.1109/TE.2005.856149

[11] Anderson, J. R., Corbett, A. T., Koedinger, K. R., & Pelletier, R. (1995). Cognitive tutors: Lessons learned. Journal of the Learning Sciences 4, 167-207 https://doi.org/10.1207/s15327809jls0402_2

[12] Kilde-Westberg, S., Johansson, A., & Enger, J. (2025). Generative AI as a lab partner: A case study. Physical Review Physics Education Research 21, 020119 https://doi.org/10.1103/ggy1-3kjk

[13] Ding, L., Li, T., Jiang, S., & Gapud, A. (2023). Students' perceptions of using ChatGPT in a physics class as a virtual tutor. International Journal of Educational Technology in Higher Education 20, 63 https://doi.org/10.1186/s41239-023-00434-1

[14] Arvidsson, K. (2025). AI-generated language in introductory physics lab reports in the first year of ChatGPT. Discover Artificial Intelligence 5, 408 https://doi.org/10.1007/s44163-025-00742-7

[15] Tufino, E., & Gregorcic, B. (2025). Creating a customisable Socratic AI physics tutor. Physics Education 60, 065037 https://doi.org/10.1088/1361-6552/ae0d23

[16] Scarlatos, A., Liu, N., Lee, J., Baraniuk, R., & Lan, A. (2025). Training LLM-based tutors to improve student learning outcomes in dialogues. Proceedings of the International Conference on Artificial Intelligence in Education, 251-266

[17] Weng, X., Xia, Q., Gu, M. Y. M., Rajaram, K., & Chiu, T. K. F. (2024). Assessment and learning outcomes for generative AI in higher education: A scoping review on current research status and trends. Australasian Journal of Educational Technology, 40(6), 37–55. https://doi.org/10.14742/ajet.9540

[18] Yan, L., Sha, L., Zhao, L., Li, Y., Martinez-Maldonado, R., Chen, G., Li, X., Jin, Y., & Gašević, D. (2024). Practical and ethical challenges of large language models in education: A systematic scoping review. British Journal of Educational Technology, 55(1), 90–112. https://doi.org/10.1111/bjet.13370

[19] Hofstein, A., & Lunetta, V. N. (2004). The laboratory in science education: Foundations for the twenty-first century. Science Education 88, 28-54 https://doi.org/10.1002/sce.10106

[20] Lunetta, V. N., Hofstein, A., & Clough, M. (2007). Learning and teaching in the school science laboratory: An analysis of research, theory, and practice. In N. G. Lederman & S. K. Abell (Eds.), Handbook of Research on Science Education (Lawrence Erlbaum), 393-441.

[21] Etkina, E., Murthy, S., & Zou, X. (2006). Using introductory labs to engage students in experimental design. American Journal of Physics 74, 979-986 https://doi.org/10.1119/1.2238885

[22] Holmes, N. G., & Wieman, C. E. (2018). Introductory physics labs: We can do better. Physics Today 71(1), 38-45 https://doi.org/10.1063/PT.3.3816

[23] Zwickl, B. M., Finkelstein, N., & Lewandowski, H. J. (2013). The process of transforming an advanced lab course: Goals, curriculum, and assessments. American Journal of Physics 81, 63-70 https://doi.org/10.1119/1.4768890

[24] Wilcox, B. R., & Lewandowski, H. J. (2016). Open-ended versus guided laboratory activities: Impact on students' beliefs about experimental physics. Physical Review Physics Education Research 12, 020132 https://doi.org/10.1103/PhysRevPhysEducRes.12.020132

[25] Kohl, P. B., & Finkelstein, N. D. (2006). Effects of representation on students solving physics problems: A fine-grained characterization. Physical Review Special Topics - Physics Education Research 2, 010106 https://doi.org/10.1103/PhysRevSTPER.2.010106

[26] Redish, E. F. (2004) A theoretical framework for physics education research: Modeling student thinking. In E. F. Redish & M. Vicentini (Eds.), Proceedings of the International School of Physics "Enrico Fermi" (IOS Press), pp. 1-63.

[27] Bing, T. J., & Redish, E. F. (2009). Analyzing problem solving using math in physics: Epistemological framing via warrants. Physical Review Special Topics - Physics Education Research 5, 020108. https://doi.org/10.1103/PhysRevSTPER.5.020108

[28] Scherr, R. E., & Hammer, D. (2009). Student behavior and epistemological framing: Examples from collaborative active-learning activities in physics. Cognition and Instruction 27, 147-174 https://doi.org/10.1080/07370000902797379

[29] Airey, J., & Linder, C. (2009). A disciplinary discourse perspective on university science learning: Achieving fluency in a critical constellation of modes. Journal of Research in Science Teaching 46, 27-49 https://doi.org/10.1002/tea.20265

[30] Bing, T. J., & Redish, E. F. (2009). Analyzing problem solving using math in physics: epistemological framing viawarrants Phys. Rev. Spec.Top.-Phys. Educ. Res. 5 020108

[31] Bing, T. J., & Redish, E. F. (2012). Epistemic complexity and the journeyman-expert transition, Phys. Rev. Spec. Top.-Phys. Educ. Res. 8 010105

[32] Thompson, J. D., Modir, B. & Sayre, E. C. (2016) Algorithmic, conceptual, and physical thinking: a framework for understanding student difficulties in quantum mechanics, Proc. Int. Conf. of the Learning Sciences

[33] Nguyen H D,Chari D N and Sayre EC 2016 Dynamics of students' epistemological framing in group problem solving Eur.J. Phys. 37 065706

[34] Fredlund, T., Linder, C., Airey, J., & Linder, A. Unpacking physics representations: Towards an appreciation of disciplinary affordance. Physical Review Special Topics - Physics Education Research 10, 020129 (2014). https://doi.org/10.1103/PhysRevSTPER.10.020129

[35] Tuveri, M., Steri, A., & Fanti, V. (2026). Semiotic problem framing: a new framework to guide students and teachers in conceptual understanding and teaching of physics, Eur. J. Phys. 47 015712, 10.1088/1361-6404/ae244

[36] Price, A., Kim, C. J., Burkholder, E., Fritz, A. V., & Wieman, C. E. (2021). An expert-decision framework for teaching and assessing problem-solving in science and engineering. International Journal of Science Education 43, 2314-2337 https://doi.org/10.1080/09500693.2021.1968334

[37] Heimbeck, M. S., Burkholder, E., & Wieman, C. E. (2025). Decision making in graduate physics coursework. Physical Review Physics Education Research 21, 010148.

[38] Barron, B. (2006). Interest and self-sustained learning as catalysts of development: A learning ecology perspective. Human Development, 49(4), 193–224. https://doi.org/10.1159/00009436

[39] Luckin, R. (2008). The learner centric ecology of resources: A framework for using technology to scaffold learning. Computers & Education, 50(2), 449–462. https://doi.org/10.1016/j.compedu.2007.09.01

[40] Drijvers, P., Doorman, L., Boon, P., Reed, H., & Gravemeijer, K. (2010). The teacher and the tool: Instrumental orchestrations in the technology-rich mathematics classroom. Educational Studies in Mathematics, 75(2), 213–234.

[41] Johri, A. (2011). The socio-materiality of learning practices and implications for the field of learning technology. Research in Learning Technology, 19(3), 207–217. https://doi.org/10.3402/rlt.v19i3.1711

[42] Johri, A. (2022). Augmented sociomateriality: Implications of artificial intelligence for the field of learning technology. Research in Learning Technology, 30, Article 2642.

[43] Hancock, J. T., Naaman, M., & Levy, K. (2020). AI-mediated communication: Definition, research agenda, and ethical considerations. Journal of Computer-Mediated Communication, 25(1), 89–100.

[44] Bailey, J. M., & Slater, T. F. (2003). A review of astronomy education research. Astronomy Education Review 2(2), 20-45

[45] Bailey, J. M., & Slater, T. F. (2005). Resource Letter AER-1: Astronomy education research. American Journal of Physics 73, 677-685

[46] Wooten, M. M., Coble, K., Puckett, A. W., & Rector, T. (2018). Investigating introductory astronomy students' perceived impacts from participation in course-based undergraduate research experiences. Physical Review Physics Education Research 14, 010151

[47] Hewitt, H. B., Simon, M. N., Mead, C., Grayson, S., Beall, G. L., Zellem, R. T., Tock, K., & Pearson, K. A. (2023). Development and assessment of a course-based undergraduate research experience for online astronomy majors. Physical Review Physics Education Research 19, 020156.

[48] Howell, S. B. (2006). Handbook of CCD Astronomy, 2nd ed. (Cambridge University Press). https://doi.org/10.1017/CBO9780511807909

[49] Birney, D. S., Gonzalez, G., & Oesper, D. (2006). Observational Astronomy, 2nd ed. (Cambridge University Press).

[50] Kolb, U., Brodeur, M., Braithwaite, N. S. J., & Minocha, S. (2018). A robotic telescope for university-level distance teaching. Robotic Telescopes, Student Research and Education Proceedings 1, 127-136

[51] OpenAI. Creating and editing GPTs. OpenAI Help Center. Retrieved June 22, 2026, from https://help.openai.com/en/articles/8554397-creating-and-editing-gpts

[52] Braun, V., & Clarke, V. Using thematic analysis in psychology. Qualitative Research in Psychology 3, 77-101 (2006). https://doi.org/10.1191/1478088706qp063oa

[53] Schreier, M. Qualitative Content Analysis in Practice (SAGE, 2012).

[54] Nowell, L. S., Norris, J. M., White, D. E., & Moules, N. J. Thematic analysis: Striving to meet the trustworthiness criteria. International Journal of Qualitative Methods 16, 1-13 (2017). https://doi.org/10.1177/1609406917733847

[55] O'Brien, B. C., Harris, I. B., Beckman, T. J., Reed, D. A., & Cook, D. A. (2014). Standards for reporting qualitative research: A synthesis of recommendations. Academic Medicine, 89(9), 1245-1251. https://doi.org/10.1097/ACM.0000000000000388

[56] Tong, A., Sainsbury, P., & Craig, J. (2007). Consolidated criteria for reporting qualitative research (COREQ): A 32-item checklist for interviews and focus groups. International Journal for Quality in Health Care, 19(6), 349-357. https://doi.org/10.1093/intqhc/mzm042

[57] Tracy, S. J. (2010). Qualitative quality: Eight big-tent criteria for excellent qualitative research. Qualitative Inquiry, 16(10), 837-851. https://doi.org/10.1177/1077800410383121

[58] Hattie, J., & Timperley, H. The power of feedback. Review of Educational Research 77, 81-112 (2007). https://doi.org/10.3102/003465430298487

[59] Nicol, D. J., & Macfarlane-Dick, D. Formative assessment and self-regulated learning: A model and seven principles of good feedback practice. Studies in Higher Education 31, 199-218 (2006). https://doi.org/10.1080/03075070600572090

## Appendix A. Post-activity reflective questionnaire

*Description: The questionnaire was anonymous and non-evaluative. Its purpose was to understand how students used AstroTutor during the laboratory. The responses were used only as limited contextual evidence.*

*Questions:*

1. In which laboratory task(s) did you use AstroTutor? Please describe the context: observation planning, use of the Observation Planner, airmass, exposure time, aperture photometry, S/N, data analysis, code or report writing.
2. Can you describe one moment in which AstroTutor helped you clarify a physical or methodological concept? Please explain what was unclear before the interaction and what became clearer afterwards.
3. Did AstroTutor help you connect a concept to an operational decision in the laboratory? For example, airmass to observing window, seeing to aperture choice, S/N to exposure time, ADU/saturation to detector limits, or comparison/check stars to differential photometry.
4. Which type of AstroTutor response was most useful for you? For example: guiding questions, step-by-step explanations, direct explanations, diagnostic feedback on your plan/report, formulas, calculations or code.
5. Were there moments in which AstroTutor was not useful, was too generic, too directive, confusing or too authoritative? Please describe one episode, if any.

6. Did you verify AstroTutor's responses against course materials, instructor indications, the Observation Planner, your data or other sources? Please describe how you checked the response.
7. Did interacting with AstroTutor change how you approached the laboratory task or the report? For example, did it help you organise reasoning, identify missing assumptions, revise a plan, justify a decision or communicate the analysis more clearly?
8. What limits or safeguards should be introduced if AstroTutor is used again in this laboratory? Please mention grading, reliability, report writing, direct answers, Socratic questions, code generation or instructor supervision.
9. Overall, what did AstroTutor support best, and what should remain the responsibility of the student and the instructor?

## Appendix B. Representative translated excerpts and analytic rationale

The excerpts below are translated from Italian and lightly edited only to remove identifying information and improve readability. They are included to make the link between empirical traces and analytic claims inspectable. They are not intended as frequency counts.

**Table B1.** Representative excerpts and analytic rationale.

| Source | Excerpt or observed move | Analytic interpretation |
|---|---|---|
| Student using planner | "My objective is to orient myself in the Observation Planner ... and write notes that recall the theoretical concepts seen in class." | Interface-interpreter framing. The student treats the software as an object of conceptual interpretation. |
| Student using planner | "If a star is close to the horizon, it has to cross a greater atmospheric thickness." | Concept-to-decision potential. Altitude is linked to atmospheric path length and therefore to airmass as an observing constraint. |
| Student using planner | "Should I also set the day of the year when I make the observation?" | The student identifies time as a hidden variable in transforming coordinates into visibility. |
| Student with observing plan | Complete V0536 Peg LaTeX plan pasted into AstroTutor for diagnostic review. | Diagnostic-reviewer and report-scaffold framing. The student brings a concrete artefact into the interaction. |
| AstroTutor | "If I had to give a laboratory grade: 28–30/30." | Role drift. The tutor crosses from formative feedback into inappropriate assessment-like authority. |
| Questionnaire user | AstroTutor was useful for clarifications, notation, formula details and code, but "only provided information" and produced "too many hallucinations" in some definitions. | Negative evidence. The student did not perceive strong concept-to-decision support and foregrounded verification burden. |